\documentclass[reprint,twocolumn,superscriptaddress,showkeys,
nofootinbib,notitlepage,amsmath,amssymb,floatfix]{revtex4-2}

\usepackage{latexsym,amsmath,amssymb,lmodern,float,url}
\usepackage{natbib}
\usepackage{color}
\usepackage{multirow}
\usepackage{bm}
\usepackage{array}
\usepackage[normalem]{ulem}
\usepackage{cleveref}
\usepackage{xcolor}

\usepackage{tikz}
\usetikzlibrary{datavisualization}
\usetikzlibrary{datavisualization.formats.functions}
\usetikzlibrary{plotmarks}

\usepackage{mathtools}

\def\de{\delta^{\vphantom{1}}}
\def\bde{{\bar\delta}}

\def\ccss{{c\bar{c}s\bar{s}}}

\def\ccqq{{c\bar{c}q\bar{q}^\prime}}

\def\h3{{\displaystyle{\frac 3 2}}}

\newcommand{\bbar}{\overline}

\def\schro{Schr\"odinger~}

\begin{document}
\title{Meson-Meson Molecular Binding Potentials in the Diabatic Dynamical Diquark Model}
\author{Richard F. Lebed}
\email{Richard.Lebed@asu.edu}
\author{Steven R. Martinez}
\email{srmart16@asu.edu}
\affiliation{Department of Physics, Arizona State University, Tempe,
AZ 85287, USA}
\date{March, 2025}

\begin{abstract}
Using the diabatic formalism, a rigorous generalization of the Born-Oppenheimer approximation, we study the effects of introducing simple meson-meson molecular potentials to an established diquark model for tetraquark states, and calculate mixed bound states composed of both diquark-antidiquark and meson-meson molecular components.  We examine the behavior of  properties of the states as one varies the parameters of the di-meson potential, and find that significant regions of parameter space occur in which one may produce mass eigenstates exactly matching the specific examples of $\chi_{c1}(3872)$ and $\chi_{c0}(3915)$.  We also find regions in which a pure di-meson molecular state emerges, and study the same state properties in such cases.
\end{abstract}

\keywords{Exotic hadrons, diquarks, scattering}
\maketitle

\section{Introduction}\label{sec:Intro}
Through the discovery of exotic hadrons, modern had\-ronic physics has entered a new era, highlighted by an incredibly deep and rich avenue of study.  The observation in 2003 of the first heavy-quark exotic hadron $X(3872)$~\cite{Choi:2003ue} [henceforth denoted by its current name $\chi_{c1}(3872)$] has led to extensive searches at multiple facilities for other candidates, which now total 70+ as of this writing.  While theoretical predictions of multiquark exotic hadrons date as far back as the introduction of the quark model by Gell-Mann~\cite{Gell-Mann:1964ewy} and Zweig~\cite{Zweig:1964ruk,Zweig:1964jf}, no single theoretical model has emerged to completely predict and describe the full suite of confirmed exotic states.

States such as $\chi_{c1}(3872)$, which lie rather close to the nearest di-meson threshold carrying the same quantum numbers, have emerged as strong candidates for identification as a molecule of those mesons.  In the case of $\chi_{c1}(3872)$, one would identify the state to be (at least mostly) comprised of a $D^0$ and a $\bar D^{*0}$ meson (plus the charge-conjugate state $\bar D^0 D^{*0}$) bound together by an attractive potential.  In fact, this concept dates back almost 30 years prior to the discovery of $\chi_{c1}(3872)$~\cite{Voloshin:1976ap,DeRujula:1975qlm}, only shortly after the development of QCD itself.  However, given the small measured mass difference~\cite{ParticleDataGroup:2024cfk} between the free meson pair $D^0 \bar D^{*0}$ and $\chi_{c1}(3872)$,
\begin{equation} \label{eq:Xbind}
m_{\chi_{c1}(3872)} - m_{D^0} - m_{D^{*0}} = -0.05 \pm 0.09 \
{\rm MeV} \, ,
\end{equation}
one expects this attractive potential to be quite shallow.  If the same dynamical mechanism is responsible for the binding of all exotic candidates, the result Eq.~(\ref{eq:Xbind}) discourages a molecular interpretation for states that lie significantly further below the nearest di-meson threshold (say, $> 10$--$15$~MeV).  Furthermore, multiple exotic candidates fail to lie immediately below a di-hadron threshold; some appear slightly above---or even quite distant from---any relevant threshold.  In such cases, naive identification as a hadronic molecule bound by light-meson exchange (the most obvious mechanism) is indeed difficult.

Additionally, $\chi_{c1}(3872)$ itself exhibits a substantial ($> \! 10\%$) decay branching fraction to charmonium, indicating that its wave function contains a significant short-range component, in stark contrast to the spatially large molecule (multiple fm) one would associate with such a small binding energy.  The simplest configuration one could associate with the short-distance component is the conventional charmonium state $\chi_{c1}(2P)$, and in fact, treating $\chi_{c1}(3872)$ as a $\chi_{c1}(2P)$-$D^0 \bar D^{*0}$ admixture has long history~\cite{Suzuki:2005ha}.  However, charmonium is not the only short-range, color-neutral configuration available to this state.  A color-antitriplet diquark $\de \! \equiv \! (cq)_{\mathbf{\bar3}}$ plus a color-triplet antidiquark $\bde \! \equiv \! (\bar c \bar q)_{\mathbf{3}}$ ($q$ being any light-quark flavor) can also satisfy this condition.  A diquark-antidiquark framework (or diquark-triquark, using an analogous color-triplet mechanism to produce pentaquarks~\cite{Lebed:2015tna}) also has the benefit of being able to accommodate exotic states that do not have a nearby relevant di-hadron threshold.  One such framework, the \textit{dynamical diquark model}~\cite{Brodsky:2014xia,Lebed:2017min}, has been thoroughly developed to produce exotic states in multiple flavor and spin sectors using a $\de$-$\bde$ paradigm~\cite{Lebed:2017min,Giron:2019bcs,Giron:2019cfc,Giron:2020fvd,Giron:2020qpb,Giron:2020wpx,Giron:2021sla,Giron:2021fnl}.

The most recent development within this model has been the incorporation of the \textit{diabatic formalism}, which allows for the inclusion of mixing between $\de \bde$ and di-meson configurations in order to produce bound~\cite{Lebed:2022vks} and scattering~\cite{Lebed:2023kbm} states.  This formalism, designed as a method to study level crossings in atomic physics~\cite{Baer:2006}, is a rigorous generalization of the Born-Oppenheimer (\textit{adiabatic}) approximation, which in turn serves as the foundation for the original formulation of the dynamical diquark model~\cite{Lebed:2017min}.  The first use of the diabatic approach for the study of heavy-quark exotic states was implemented similarly, by introducing mixing between quarkonium and di-meson thresholds~\cite{Bruschini:2020voj}.

While previous dynamical-diquark works have focused solely on using the diabatic formalism as a method for coupling $\de\bde$ configurations to di-meson thresholds ({\it i.e.}, free di-meson states), it is also possible to use this framework to examine mixing between diquark and bound molecular pictures.  The simplest way to accomplish such studies, as foreshadowed several times in previous works~\cite{Lebed:2022vks,Lebed:2023kbm,Lebed:2024rsi}, is to introduce a shallow, attractive di-meson potential into the calculations in place of the trivial free meson-pair interaction term previously used.  

In this work we examine the effects of introducing this potential by varying its shape, parameters, and the specific exotic states to which it is applied. We find that its effect is largely as might be anticipated: In the limit that the attractive potential is shallow, the mass eigenvalue of the full state shifts downward, but the state persists.  If the potential becomes sufficiently deep, the system develops a numerical instability in the form of a new bound state, which one may interpret as a pure di-meson molecular state, as opposed to the original $\de\bde$-di-meson composite.  In most of our calculations, one takes the mass eigenvalue as fixed by experiment, in which case either the diquark mass or the diquark fractional content of the full state must be adjusted to accommodate the known mass.  In all cases, our detailed numerical results support simple phenomenological explanations, and serve to provide specific benchmarks for understanding the scaling behaviors of the diabatic dynamical diquark model.

This paper is structured as follows.  In Sec.~\ref{sec:model} we briefly review the incorporation of the diabatic formalism into the dynamical diquark model.  In Sec.~\ref{sec:results} we present the results of including a shallow, attractive potential between the di-meson pairs of a $J^{PC}=1^{++}$ hidden-charm state [relevant to $\chi_{c1}(3872)$] and a $0^{++}$ hidden-charm hidden-strange state [relevant to $\chi_{c0}(3915)$].  We then summarize our findings and describe future prospects of this model in Sec.~\ref{sec:concl}.

\section{The Diabatic Dynamical Diquark Model}\label{sec:model}

The dynamical diquark \textit{picture}~\cite{Brodsky:2014xia,Lebed:2015tna} describes the formation of a diquark-antidiquark pair formed with high relative momenta such that they quickly separate, transferring their available kinetic energy into the color flux tube connecting them.  This resulting configuration will continue to have difficulty hadronizing due to the separation of quarks from antiquarks, thus allowing it to persist long enough to be measured as an exotic state.  The dynamical diquark \textit{model}~\cite{Lebed:2017min} then uses the Born-Oppenheimer (BO) approximation to extract a spectrum of static bound states from this quasi-static configuration.\footnote{Key in the development of the model is the expectation that each diquark quasiparticle is suitably compact (a radius of a few times 0.1~fm), due to its nucleation about the heavy constituent quark, while the flux tube connecting them is at least somewhat larger.  This hypothesis has been tested numerically in Ref.~\cite{Giron:2019cfc} and is found not to severely disrupt the predicted tetraquark mass spectrum, even for diquarks of radius as large as 0.7~fm\@.  This underlying structure is reminiscent of the ``butterfly'' 4-quark configuration~\cite{Vijande:2007ix} of color-string models.  In the opposite limit (as also discussed in Ref.~\cite{Giron:2019cfc}), in which the heavy quark-antiquark pair separation becomes smaller than the typical diquark size, the configuration becomes one resembling that of the \textit{hadroquarkonium} picture~\cite{Dubynskiy:2008mq}, in which a heavy quarkonium state is surrounded by a light-quark cloud.}.  Here we briefly describe this model, and how one transitions to the rigorous BO generalization known as the \textit{diabatic formalism}.   One first separates the heavy and light degrees of freedom (d.o.f.):
\begin{equation} 
\label{eq:SepHam}
H=K_{\rm heavy} + H_{\rm light} =
\frac{\mathbf{p}^2}{2 \mu_{\rm heavy}} + H_{\rm light},
\end{equation}
so that one may apply the adiabatic (conventional BO) expansion to the overall solution:
\begin{equation} 
\label{eq:AdExp}
|\psi \rangle = \sum_{i} \int d\mathbf{r} \, \tilde \psi_i
(\mathbf{r}) \, |\mathbf{r} \rangle \:
|\xi_i(\mathbf{r}) \rangle \, ,
\end{equation}
where $|\mathbf{r} \rangle$ is the position-space eigenstate denoting the separation $\mathbf{r}$ between the heavy color sources of the system, $|\xi_i(\mathbf{r}) \rangle$ are the eigenstates of $H_{\rm light}$ for a given value of $\mathbf{r}$, and $\tilde \psi_i (\mathbf{r})$ is the $\mathbf{r}$-dependent amplitude for the \textit{i}$^{\rm th}$ component.  Physically, this expansion describes a state for which the light d.o.f.\ \textit{instantaneously} adapt to changes in the separation $\mathbf{r}$ of the heavy color sources.  Inserting Eq.~(\ref{eq:AdExp}) into the \schro equation and using the Hamiltonian of Eq.~(\ref{eq:SepHam}) gives
\begin{equation}
\sum_i \left( - \frac{\hbar^2}{2 \mu_i} [\mathbf{\nabla} + \tau (\mathbf{r})]^2_{ji} + [V_j (\mathbf{r}) - E] \, \delta_{ji}  \right) \! \tilde \psi_i (\mathbf{r}) = 0,
\end{equation}
where we have adopted the definition~\cite{Baer:2006}
\begin{equation} \label{eq:NACT}
\mathbf{\tau}_{ji}(\mathbf{r}) \equiv \langle \xi_j (\mathbf{r}) | \nabla \xi_i (\mathbf{r}) \rangle,
\end{equation}
thus defining the \textit{non-adiabatic coupling terms} (NACTs) $\tau_{ji}$.  Under the BO approximation, one then implements the \textit{single-channel approximation}, \textit{i.e.},
\begin{equation}
\tau_{ji} (\mathbf{r}) = \langle \xi_j (\mathbf{r}) | \nabla \xi_i (\mathbf{r}) \rangle \simeq 0.
\end{equation}
This assumption then yields a much simpler, decoupled set of \schro equations, the hallmark of the BO approach.

However, the energy of this system may naturally lie close to the threshold for creating a free di-meson pair $M\bar M^\prime$; in particular, the $\de \bde$ and $M\bar M^\prime$ systems possess the same quark content, and therefore through the sum of their constituent masses have equal energy at lowest order.  As discussed in the Introduction, a mechanism to incorporate these thresholds is imperative for creating a complete model of exotic hadrons, as numerous candidates have experimentally measured mass eigenvalues that match this description.  The \textit{diabatic formalism}, which serves as a rigorous generalization of the BO approximation, allows for the explicit incorporation of di-meson thresholds into the dynamical diquark model.  Here we briefly describe the application of this formalism.

In order to include di-meson threshold mixing appro\-priately, one must not only relax the single-channel approximation, but also reexamine the expansion of Eq.~(\ref{eq:AdExp}) to allow for the accommodation of \textit{diabatic} (level-crossing) effects in the system near di-meson thresholds.  Under the diabatic formalism, Eq.~(\ref{eq:AdExp}) is replaced with 
\begin{equation} 
\label{eq:DiaExp}
|\psi \rangle = \sum_{i} \int d\mathbf{r}' \tilde \psi_i
(\mathbf{r}' \! , \mathbf{r}_0) \: |\mathbf{r}' \rangle \:
|\xi_i(\mathbf{r}_0) \rangle,
\end{equation}
where the expansion is performed using a particular arbitrary fiducial separation $\mathbf{r}_0$.  Of course, this expansion is equally as valid as Eq.~(\ref{eq:AdExp}), since the set of states $|\xi_i(\mathbf{r}_0) \rangle$ is complete for all values of $\mathbf{r}_0$, but the introduction of distinct separation parameters $\mathbf{r}^\prime$ and $\mathbf{r}_0$ in $\tilde \psi_i
(\mathbf{r}' \! , \mathbf{r}_0)$ provides a natural parametrization for non-instantaneous couplings between the light and heavy d.o.f.\ of the system.  Reinserting this expansion into the \schro equation defined by Eq.~(\ref{eq:SepHam}) now gives
\begin{equation}\label{eq:DiaSchro}
\sum_{i} \left[ - \frac{\hbar^2}{2 \mu_{i}} \de_{ij}  \nabla ^2 +
V_{ji}(\mathbf{r,r_0})-E \de_{ji} \right] \! \tilde \psi_i (\mathbf{r,r_0}) = 0,
\end{equation}
where the \textit{diabatic potential matrix} $V_{ji}$~\cite{Bruschini:2020voj}, rigorously equivalent~\cite{Baer:2006} to using NACTs [Eq.~(\ref{eq:NACT})], is defined by
\begin{equation} \label{eq:DiaV}
V_{ji}(\mathbf{r,r_0}) \equiv \langle \xi_j (\mathbf{r}_0)|
H_{\rm light} |\xi_i(\mathbf{r}_0) \rangle.
\end{equation}
At this point, we utilize our choice of $\mathbf{r}_0$; an appropriately chosen separation $\mathbf{r}_0$ far from level crossings of unmixed states allows for the identification of the light-field eigenstates with pure diquark-antidiquark or pure di-meson configurations (See Ref.~\cite{Lebed:2022vks} for a more detailed discussion).  Thus, one may write the potential matrix as  
\begin{equation} \label{eq:FullV}
\text V=
\begin{pmatrix}
V_{\de \bde}(\mathbf{r}) & V_{\rm mix}^{(1)}(\mathbf{r})  & \cdots &
V_{\rm mix \vphantom{\bbar M_2}}^{(N)}(\mathbf{r}) \\
V_{\rm mix}^{(1)}(\mathbf{r}) & 
V_{M_1 \bbar M_2}^{(1)}(\mathbf r) &
&
\\
\vdots
& & \ddots \\
V_{\rm mix \vphantom{\bbar M_2}}^{(N)}(\mathbf{r}) & & &
V_{M_1 \bbar M_2}^{(N)}(\mathbf r) \\
\end{pmatrix},
\end{equation}
where the elements left blank are taken to be zero: \textit{i.e.}, we ignore direct mixing between di-meson thresholds.  This assumption is not necessary in order to perform a successful analysis of the system, but it is a typical ansatz for calculations of this form~\cite{Baer:2006}.  The diagonal elements are the static light-field energies associated with their corresponding unmixed configurations. 
In particular, the element $\text V_{11}$ represents the BO potentials of the $\de \bde$ configuration.  For the di-meson configurations, our prior works~\cite{Lebed:2022vks,Lebed:2023kbm,Lebed:2024rsi,Lebed:2024zrp} previously set
\begin{equation}
V_{M_1 \bbar M_2}^{(i)}(\mathbf r) \to T_{M_1 \bbar M_2} \equiv M_1 + M_2 \, ,
\end{equation}
so that the potential corresponding to each di-meson pair [labeled by $(i)$] deemed sufficiently close to the unmixed $\de\bde$ configuration energy $V_{\de\bde} (\mathbf{r})$ is given directly by its free energy (\textit{i.e.}, threshold).  However, in this work we introduce a shallow, attractive potential to $V_{M_1 \bbar M_2}^{(i)}(\mathbf r)$, to more accurately represent the true dynamics of the di-meson pair.  This modification is straightforward to incorporate numerically once the diabatic formalism has been implemented, but it also represents an important conceptual step forward by creating a close connection between diquark models and pure di-meson molecular models.

\section{Results}\label{sec:results}

The simplest interaction one can devise to respect the expected di-meson dynamics (attractive at small separation ${\bf r}$, finite range) is an attractive square-well potential,
\begin{equation} \label{eq:Vsquare}
    V_{M_1 \bbar M_2}^{(i)}(\mathbf r)=
    \begin{cases}
        T_{M_1 \bbar M_2} - V_0, &  r \leq r_c \\
        T_{M_1 \bbar M_2}, & r > r_c
    \end{cases},
\end{equation}
with the cutoff range $r_c$ [expected from typical hadronic dynamics to be $O({\rm fm})$] and the potential-well depth $V_0$ at the origin [expected to be $O({\rm MeV})$], initially being free parameters of the system.  Of course, a square-well potential represents just one idealized choice.  In order to obtain robust results confirming that the essential physics of di-meson attraction has been incorporated into the calculation, other plausible potential profiles must be explored.  We therefore also analyze the case of a truncated simple harmonic oscillator (SHO) potential,
\begin{equation} \label{eq:VHO}
    V_{M_1 \bbar M_2}^{(i)}(\mathbf r) =
    \begin{cases}
	  T_{M_1 \bbar M_2} + \frac{1}{2} k_{M \bbar M} (r^2 - r^{2}_{c}), 
	  & r \leq r_c \\
	  T_{M_1 \bbar M_2}, & r > r_c
    \end{cases} ,
\end{equation}
where we take the spring constant $k_{M \bbar M}$ to be the same for all di-meson pairs (henceforth abbreviated as $k$).  Such a potential is arguably more physical than the square well because one expects the di-meson interaction to be largest at small separation.  Detailed studies such as in  Ref.~\cite{Vijande:2009kj}, and even much earlier in Ref.~\cite{Swanson:1992ec}, lend support to this assessment.  Implementing any other potential $V_{M_1 \bbar M_2} (\textbf{r})$ respecting these basic features is expected to lead to results similar to those presented below.

Using either Eq.~(\ref{eq:Vsquare}) or (\ref{eq:VHO}) introduces two additional free parameters [($V_0$, $r_c$) or ($k$, $r_c$), respectively] into the calculation of bound states [Eqs.~(\ref{eq:DiaSchro})-(\ref{eq:DiaV})] via the 22, 33, $\ldots$, $NN$ elements of Eq.~(\ref{eq:FullV}).  We retain the same BO $\Sigma^{+}_g$ potential for V$_{11} = V_{\de \bde}(r)$ as used in previous diabatic dynamical diquark model calculations~\cite{Lebed:2022vks,Lebed:2023kbm,Lebed:2024rsi}:
\begin{equation}
\label{eq:sgmapot}
V_{\de \bde}(r)=- \frac{\alpha}{r} + \sigma r + \bar V_0 + m_{\de} +
m_{\bde} \, ,
\end{equation}
where $\alpha,\sigma,$ and $\bar V_0$ are given by~\cite{Morningstar:2019}
\begin{eqnarray} \label{eq:sgmapotparams}
\alpha & = & 0.053 \ \rm{GeV} \! \cdot \rm{fm}, \nonumber \\
\sigma & = & 1.097 \ \rm{GeV \! /fm}, \nonumber \\
\bar V_0 & = & -0.380 \ \rm{GeV}.
\end{eqnarray}
We also use the same mixing potential~\cite{Lebed:2022vks,Bruschini:2020voj}:
\begin{equation} \label{eq:Mixpot}
|V_{\rm mix}^{(i)} (r)| = \frac{\Delta}{2}
\exp \! \left\{ -\frac 1 2 \frac{\left[
V^{\vphantom\dagger}_{\de \bde}(r) -
T_{M_1 \bbar M_2 }^{(i)} \right]^2}{(\sigma \rho)^2} \right\} ,
\end{equation}
where $\rho, \, \Delta$ are given by~\cite{Lebed:2023kbm}\footnote{This form, which assumes universal values of $\rho , \Delta$ for all di-meson pairs $M_1 \bbar M_2$, ignores important dynamics related to the specific quantum numbers of the state $M_1 \bbar M_2$, particularly its spin.  These effects can be explicitly taken into account in such a way as to respect heavy-quark spin symmetry~\cite{Bruschini:2023zkb}.}
\begin{equation} \label{eq:Mxpotparams}
\rho = 0.165 ~\rm{fm}, \ \ 
\Delta = 0.295 ~\rm{GeV}.
\end{equation}
We use these same numerical values of parameters both in order to reduce the number of d.o.f.\ within the parameter space, and for consistency of comparison with prior results.  Thus, only three free parameters remain: $m_{\de} (= \! m_{\bde})$, $k$ (or $V_0$ for the square well), and $r_c$.  Note that in Eq.~(\ref{eq:VHO}) one could replace either $k$ or $r_c$ with the depth of the di-meson potential at $r \! = \! 0$, since $V_{M_1 \bbar M_2}^{(i)}(\mathbf r=0) \equiv -V_0 =-\frac{1}{2} k_{M \bbar M} \, r^{2}_{c}$; this identification is especially useful in framing the scale of the $(k,r_c)$ space, which is generally explored in regions where $V_{M_1 \bbar M_2}^{(i)}(\mathbf r \! = \! 0) \agt -20$~MeV: {\it i.e.}, not much more attractive than the typical binding scale $O(-10~{\rm MeV})$ of a nucleon within a nucleus.  For ease of comparison between the two potentials, in the square-well case we define an effective value $k \equiv 2V_0/r_c^2$.

Of course, since the full model contains two more free parameters than explored here (namely, $\Delta$ and $\rho$), one can seek additional physical input to constrain the full set.  While the literature contains some useful phenomenological constraints on $\Delta$ and $\rho$ ({\it e.g.}, Refs.~\cite{Lebed:2022vks,Bruschini:2020voj}), as well as various determinations of values for $m_{\de}$ ({\it e.g.}, Ref.~\cite{Kleiv:2013dta}), future work could focus upon possible methods for obtaining \textit{ab initio} constraints on the di-meson potential parameters $r_c$ and $k$.  One plausible method for doing so lies in examining observables related to the range and strength of the di-meson interaction included in the diabatic potential matrix, such as observed spatial correlations between any of the meson-meson configurations to which the exotic state decays, a technique known as \textit{femtoscopy}.  Recent experimental work is already addressing similar scenarios; for example, Ref.~\cite{ALICE:2024bhk} employs femtoscopic techniques to obtain correlation functions between charmed and light-flavor meson pairs.

While one could certainly analyze the effects of the di-meson potential on the full suite of mass eigenstates previously found in Refs.~\cite{Lebed:2022vks,Lebed:2023kbm}, we restrict the calculations presented here to the states associated with $\chi_{c1}(3872)$ and $\chi_{c0}(3915)$ [formerly $X(3872)$ and $X(3915)$, respectively].  These particular examples are chosen partly for generic illustrative reasons, and partly because they are special: As noted in Eq.~(\ref{eq:Xbind}), $\chi_{c1}(3872)$ has an extremely small binding energy, while $\chi_{c0}(3915)$ has been argued since Ref.~\cite{Lebed:2016yvr} to represent the lightest of the $c\bar c s\bar s$ exotic states.  Despite these distinctive properties, numerical results for other exotic states should be entirely analogous.

We thus explore fits to the 3-dimensional parameter space ($m_{\de (\bde)}$, $V_0$, $r_c$) or ($m_{\de (\bde)}$, $k$, $r_c$), constrained by the experimentally measured masses of these states~\cite{ParticleDataGroup:2024cfk}:
\begin{equation} \label{eq:ExpMasses}
\begin{split}
M_{\chi_{c1}(3872)} &= 3.87164 \pm 0.00006 \ {\rm GeV}, \\
M_{\chi_{c0}(3915)} &= 3.9221 \ \,  \pm 0.0018 \ {\rm GeV} .
\end{split}
\end{equation}
All figures and tables presented here are generated by a similar procedure: one of the three parameters (typically the cutoff $r_c$) is fixed \textit{a priori}, and the remaining two are left as free parameters in the fitting algorithm, in order to study their parametric dependence in the model.  Let us begin with the SHO potential, Eq.~(\ref{eq:VHO}).

For both states [$\chi_{c1}(3872)$ or $\chi_{c0}(3915)$], we find that any combination of $(k,r_c)$ in Eq.~(\ref{eq:VHO}) producing a potential-well depth $V_0 < 18$~MeV allows for an exact fit to Eqs.~(\ref{eq:ExpMasses}).  Explicitly, the absolute value of the difference between the calculated and experimental mass, $|M-M_X|$ [where $M_X$ is either value in Eqs.~(\ref{eq:ExpMasses})], can always be kept within $0.1$~keV\@.  We note that this result is obtained using a maximum value of $r_c=1.9$~fm, which is relatively small compared to some expectations for the size of exotic molecules---particularly $\chi_{c1}(3872)$---as discussed below.  The region of parameter space with $V_0 < 18$~MeV (and $r_c \leq 1.9$ fm) then specifies the domain in which one has the most freedom to manipulate the other two parameters while still maintaining a perfect fit to Eqs.~(\ref{eq:ExpMasses}).  The parameter space outside of this region, which we discuss below, contains regions in which one cannot always maintain a good fit to Eqs.~(\ref{eq:ExpMasses}).

\begin{figure} 
    \centering
    \includegraphics[scale=0.5]{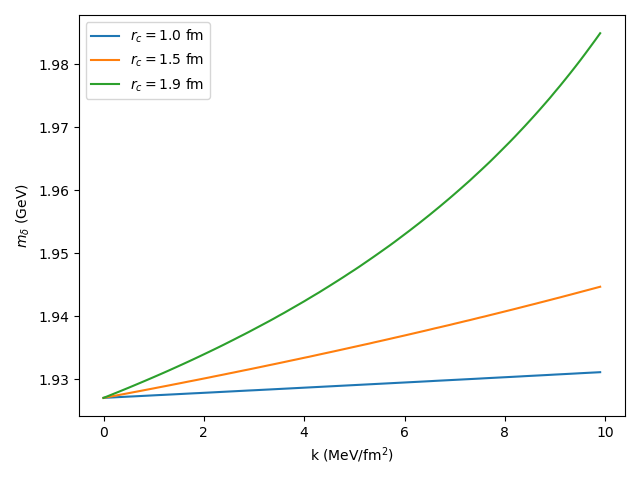}
    \caption{The $\ccqq$ diquark mass $m_{\de =(cq)}$ as a function of $k$ for various values of $r_c$ (increasing from bottom to top), using the SHO potential Eq.~(\ref{eq:VHO}).}
    \label{fig:1++_k_mde_fine}
\end{figure}

\begin{figure} 
    \centering
    \includegraphics[scale=0.5]{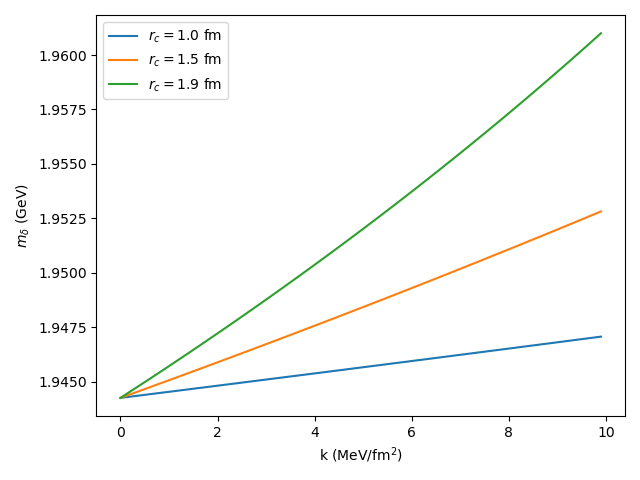}
    \caption{The $\ccss$ diquark mass $m_{\de =(cs)}$ as a function of $k$ for various values of $r_c$ (increasing from bottom to top), using the SHO potential Eq.~(\ref{eq:VHO}).}
    \label{fig:0++_k_mde_fine}
\end{figure}

Furthermore, for both states we find that the diquark mass $m_{\de}$ must always be increased with $k$ in order to maintain the fit to Eqs.~(\ref{eq:ExpMasses}), given a fixed value of $r_c$: As one might expect, deeper potentials require greater $m_\de$ values to counteract their effect and preserve the fixed mass eigenvalue.  We also find the second derivative of $m_\de$ {\it vs.}\ $k$ to vary exponentially with the potential cutoff $r_c$, as can be determined from Figs.~\ref{fig:1++_k_mde_fine}--\ref{fig:0++_k_mde_fine}:\footnote{The surprisingly small difference between $m_{(cq)}$ and $m_{(cq)}$ apparent from Figs.~\ref{fig:1++_k_mde_fine}--\ref{fig:0++_k_mde_fine} has been discussed at length in Ref.~\cite{Lebed:2024rsi}.  Ultimately, it arises from the assertion that $\chi_{c0}(3915)$, a putative $c\bar c s\bar s$ state, lies only slightly above $\chi_{c1}(3872)$ in mass.} Specifically, starting with the $c\bar c q\bar q$ $1^{++}$ state, if 
\begin{equation} \label{eq:mdevsksquare}
    m_{\de={(cq)}} = ak^2+bk+c,
\end{equation}
with $k$ in units of MeV/fm$^2$, then
\begin{equation} \label{eq:2ndderivvsrc}
    a=Ae^{Br_c},
\end{equation}
where $A=4.3~\rm MeV^{-1}~fm^{4}$ and $B=5.962~\rm fm^{-1}$.  Similarly, for the $0^{++}$ $c\bar c s\bar s$ state, we find $A=19.7~\rm MeV^{-1}~fm^{4}$ and $B=3.824~\rm fm^{-1}$.   While this exponential dependence represents a surprising level of sensitivity to the cutoff parameter $r_c$, we will identify its source below, after discussing our results for the square well potential.

\begin{figure} 
    \centering
    \includegraphics[scale=0.5]{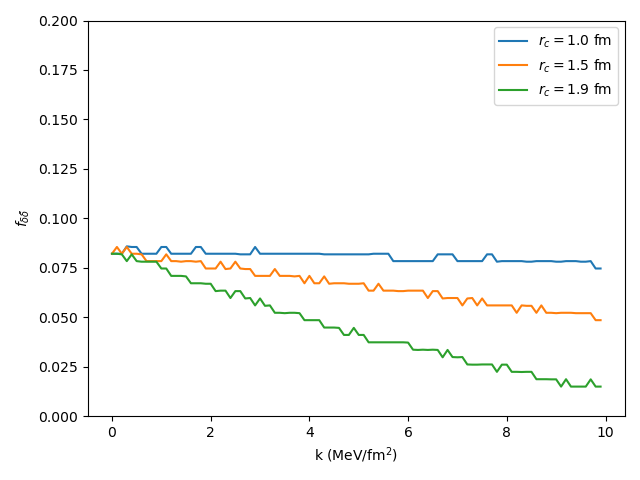}
    \caption{The $\de \bde$ fractional state content $f_{\de \bde}$ as a function of $k$ for various values of $r_c$ (increasing from top to bottom) for the $\ccqq ~ 1^{++}$ state, using the SHO potential Eq.~(\ref{eq:VHO}).  Discontinuities in the fit curve indicate that the extraction of this parameter lies at the edge of numerical precision for our algorithm.}
    \label{fig:1++_k_content_fine}
\end{figure}

\begin{figure} 
    \centering
    \includegraphics[scale=0.5]{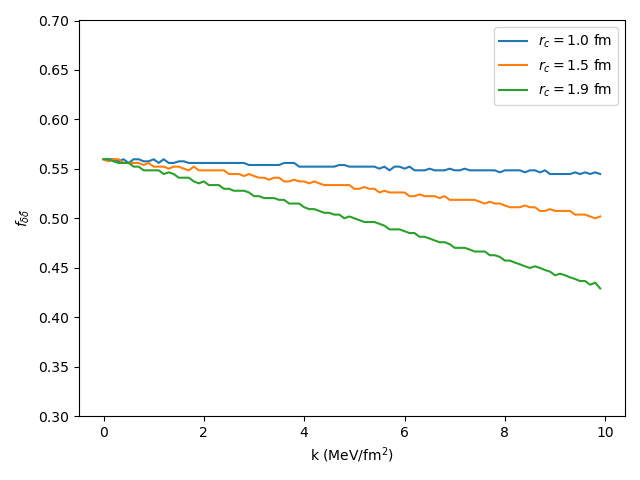}
    \caption{The $\de \bde$ fractional state content $f_{\de \bde}$ as a function of $k$ for various values of $r_c$ (increasing from top to bottom) for the $\ccss ~ 0^{++}$ state, using the SHO potential Eq.~(\ref{eq:VHO}).  Discontinuities in the fit curve indicate that the extraction of this parameter lies at the edge of numerical precision for our algorithm.}
    \label{fig:0++_k_content_fine}
\end{figure}

\begin{figure} 
    \centering
    \includegraphics[scale=0.5]{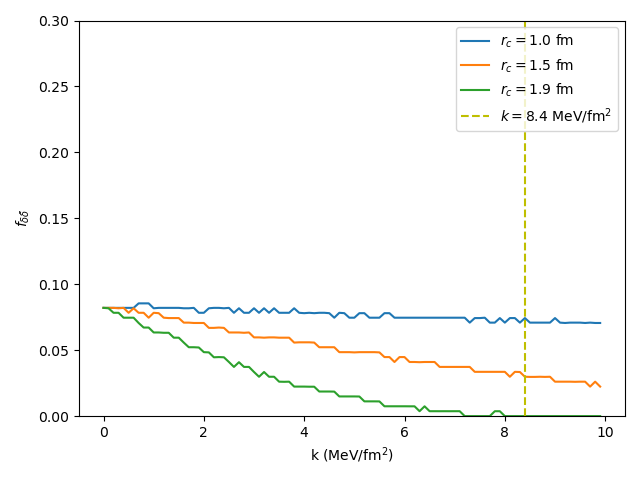}
    \caption{The $\de \bde$ fractional state content $f_{\de \bde}$ as a function of $k$ for various values of $r_c$ (increasing from top to bottom) for the $\ccqq ~ 1^{++}$ state, using the square well potential Eq.~(\ref{eq:Vsquare}).  Discontinuities in the fit curve indicate that the extraction of this parameter lies at the edge of numerical precision for our algorithm.}
    \label{fig:1++_k_content_fine_SP}
\end{figure}

In addition, we find (Figs.~\ref{fig:1++_k_content_fine}-\ref{fig:0++_k_content_fine}) that the $\de \bde$ fractional state content $f_{\de\bde}$ of the full mixed state universally decreases with $k$, once again with quadratic $k$ dependence.  As expected, $k = 0$ reproduces our previous $f_{\de\bde}$ results~\cite{Lebed:2023kbm} for both states.  Across all SHO potential simulations, the largest total drop in $f_{\de \bde}$ in absolute percentage over the range $k = 0$--$10$~MeV/fm$^2$ is $-13.1\%$ in the $0^{++}$ state (see Fig.~\ref{fig:0++_k_content_fine}).  Explicitly, for any given $r_c \leq 1.9$~fm, one may write
\begin{equation} \label{eq:fdequad}
    f_{\de \bde} = \alpha k^2 + \beta k + \gamma, 
\end{equation}
where once again the fit parameters themselves are dependent upon $r_c$.  We note that here, as opposed to the $m_{\de}$ case, the coefficient $\alpha$ is not nearly as sensitive to $r_c$, even in the case in which the pure di-meson state emerges along the curve ($r_c=1.9$ fm in Fig.~\ref{fig:1++_k_content_fine_SP} for the square well potential, to be discussed later).  This result indicates that the model produces a relatively stable state admixture ($\de\bde$ {\it vs.}\ di-meson) when one varies combinations of the di-meson potential parameters.  Here we avoid a detailed quantitative discussion of the parametric dependences of the $f_{\de \bde}(k)$ coefficients, both for brevity and for sake of maintaining the generality of the results; we merely state that we find $\alpha$ ($\beta$) to be best fit to a quadratic (linear) dependence upon $r_c$.

\begin{figure} 
    \centering
    \includegraphics[scale=0.5]{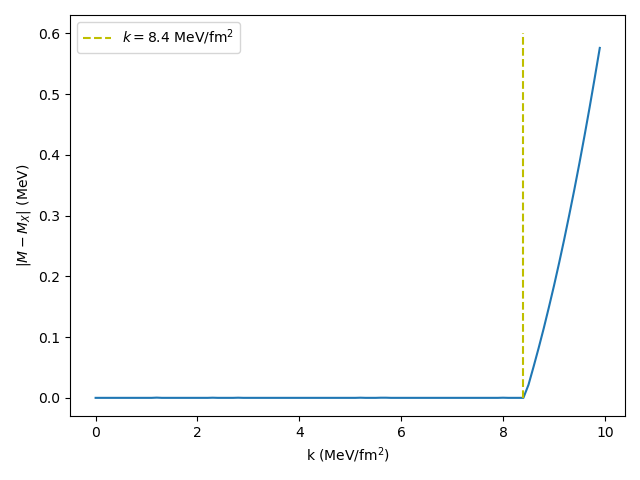}
    \caption{The $\ccqq$ $1^{++}$ mass difference $|M - M_X|$ as a function of $k$ for $r_c=1.9$ fm, using the square well potential Eq.~(\ref{eq:Vsquare}).}
    \label{fig:1++_k_massdiff_190_fine_SP}
\end{figure}

\begin{figure} 
    \centering
    \includegraphics[scale=0.5]{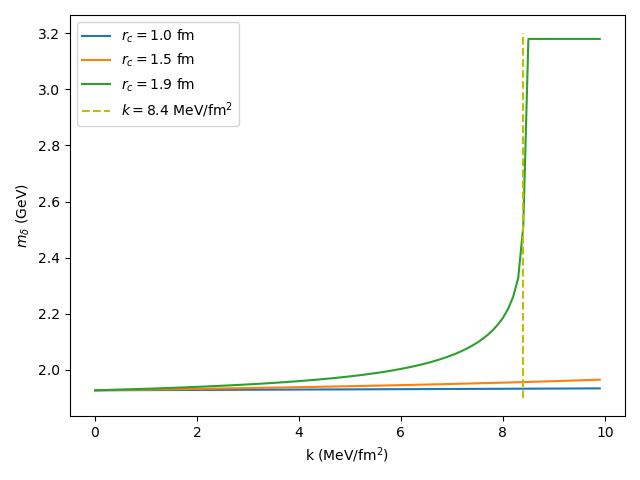}
    \caption{The $\ccqq$ diquark mass $m_{\de =(cq)}$ as a function of $k$ for various values of $r_c$ (increasing from bottom to top), using the square well potential Eq.~(\ref{eq:Vsquare}).}
    \label{fig:1++_k_m_de_fine_SP}
\end{figure}

Comparing these SHO potential results to those using the square well potential Eq.~(\ref{eq:Vsquare}), we find relatively similar behavior across all parameter space.  Perhaps most notable among the differences is that the square well potential singles out some $(k,r_c)$ pairs that do not allow a precise fit to Eqs.~(\ref{eq:ExpMasses}), but for which the fit succeeds using the SHO potential.  As mentioned in the Introduction, this inability to achieve a perfect fit is ascribed to the emergence of a distinct, pure molecular bound state. The effect is clearly depicted in Fig.~\ref{fig:1++_k_massdiff_190_fine_SP}, in which the best-fit mass difference $|M-M_X|$ abruptly begins rising to values much larger than $0.1$~keV\@.  Interestingly, under the present parameter ranges, this instability only occurs for the $1^{++}$ state, and then only for the largest value of $r_c=1.9~\rm fm$.  The dramatic increase in $|M-M_X|$ begins at $k=8.4~\rm MeV/fm^2$, corresponding to a potential well depth of $V_0 = 15.2$ MeV\@.  The corresponding results for $m_{\de\bde}$, as seen in Fig.~\ref{fig:1++_k_m_de_fine_SP}, show an abrupt plateau at the same critical $k$ value, reiterating that neither adjustments to $m_{\de}$ nor to $f_{\de\bde}$ can maintain the fixed values of Eqs.~(\ref{eq:ExpMasses}), and thus this instability indicates the generation of a new mass eigenvalue.  We note that the maximum $m_{\de}$ value appearing in Fig.~\ref{fig:1++_k_m_de_fine_SP} is quite large, nearing double its value for $k=0$.  The last, and perhaps most direct, demonstration of this effect is presented in Fig.~\ref{fig:1++_k_content_fine_SP}, in which we see that $f_{\de \bde} \to 0$ at $k=8.4~\rm MeV/fm^2$: At this point, the resulting mass eigenstate thus becomes purely di-meson molecular in content.

Returning to the space of square-well potential parameters that {\em do} allow an exact fit to Eqs.~(\ref{eq:ExpMasses}), we perform fits analogous to those discussed above in Eqs.~(\ref{eq:mdevsksquare})--(\ref{eq:2ndderivvsrc}), and discuss the resulting parameters here.  Notably, the same functional fit form works for either potential: The coefficient $a$ of $m_{\de}(k)$ in Eq.~(\ref{eq:mdevsksquare})  remains exponentially dependent upon $r_c$ [Eq.~(\ref{eq:2ndderivvsrc})], with now $A=3.2~\rm MeV^{-1}~fm^{4}$ and $B=7.292~\rm fm^{-1}$ for the $1^{++}$ state, and $A=3.6~\rm MeV^{-1}~fm^{4}$ and $B=4.030~\rm fm^{-1}$ for the $0^{++}$ state. As was true in the SHO case (Figs.~\ref{fig:1++_k_content_fine}-\ref{fig:0++_k_content_fine}), $f_{\de \bde}$ for the square well potential continues to decrease quadratically with increasing $k$, as in Eq.~(\ref{eq:fdequad}). 

Let us now examine the origin of the exponential dependence of $m_\de (k)$ indicated by Eq.~(\ref{eq:2ndderivvsrc}). As seen in Fig.~\ref{fig:1++_k_m_de_fine_SP} in the $m_{\de}$ {\it vs.}\ $k$ curves (for the square-well case), the emergence of the pure di-meson state at critical values of $(k, r_c)$ forces the fitting algorithm to rapidly increase $m_{\de}$ values in the vicinity of these critical points, in order to maintain Eqs.~(\ref{eq:ExpMasses}).  Since this transition is highly sensitive to changes in $r_c$ [see, {\it e.g.}, Eq.~(\ref{eq:kcritSquare}) below], then one expects $m_{\de}(k)$ also to be highly sensitive to $r_c$ as one approaches the critical points.  We remind the reader that other parameters of this model [Eqs.~(\ref{eq:sgmapotparams}) and (\ref{eq:Mxpotparams})] remain fixed in these fits; if they were allowed to vary, the changes in $m_{\de}$ fit values should be less drastic as one approaches points at which the pure di-meson state is generated.

As we have seen, the di-meson potential parameter space clearly possesses regions that result in the emergence of a pure molecular bound state.  Leaving $m_{\de}$ as a free parameter enables the generation of this state if one increases either the range of $k$ or of $r_c$, and since the physical implications of these changes are vastly different, we explore both below.

\begin{figure} 
    \centering
    \includegraphics[scale=0.5]{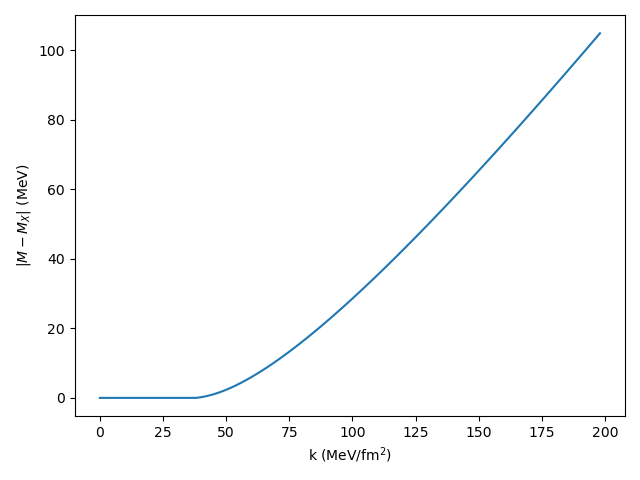}
    \caption{The $\ccqq$ $1^{++}$ mass difference $|M - M_X|$ as a function of $k$ for $r_c=1.5$ fm, using the SHO potential Eq.~(\ref{eq:VHO}).}
    \label{fig:1++_k_massdiff_Cutoff_150}
\end{figure}

Allowing for unphysically large $k$ values, even up to $k_{M\bbar M}= 200$~MeV/fm$^2$, we find that for small $r_c$ ($\leq1.0$~fm) one may always obtain a physically acceptable value of $m_{\de}$ that reproduces Eqs.~(\ref{eq:ExpMasses}).  However, in the previous search region of $r_c= 1.0$--$1.9$~fm, the emergence of the new mass eigenvalue nearly always occurs, especially for $k$ in the range $100$--$200$~MeV/fm$^2$.  An explicit example is exhibited in Fig.~\ref{fig:1++_k_massdiff_Cutoff_150}, which depicts, for the SHO potential, the analogous $|M-M_X|$ behavior to that seen in Fig.~\ref{fig:1++_k_massdiff_190_fine_SP} for the square well potential.  However, the resulting maximum difference is much larger, reaching just over $100~ \rm MeV$\@. Before continuing to examine this numerical instability in more detail, we again remind the reader that these specific values must be interpreted only semi-quantitatively, as several other parameters of the model [in particular, Eqs.~(\ref{eq:sgmapotparams}) and (\ref{eq:Mxpotparams})] that are fixed at this stage might still be adjusted to achieve an exact fit to Eqs.~(\ref{eq:ExpMasses}).  However, at some point these values might become so unphysically large that the model ceases to provide a meaningful description of the states.  For a sufficiently strong potential, the emergence of pure di-meson molecular states independent of $\de\bde$ mixing should be inevitable.

Probing the $(k,r_c)$ SHO potential parameter space to find the range of values at which $|M-M_X| \leq 0.1$~keV can no longer be achieved, we obtain a power-law relationship:
\begin{equation} \label{eq:kcritHO}
    k^{\textrm{SHO}}_{\rm crit}=0.177 \cdot r^{-3.743}_{c},
\end{equation}
which means that $k^{\textrm{SHO}}_{\rm crit}$ (in GeV/fm$^2$) is the smallest value of $k$ for a given value of $r_c$ (in fm) at which fits to Eqs.~(\ref{eq:ExpMasses}) become unstable.  This result provides a first quantitative determination of constraints on specific potential parameters that allow for a smooth introduction of the di-meson potential into the existing structure of the diabatic dynamical diquark model.  For stronger potentials, additional pure di-meson molecular states arise.

\begin{figure} 
    \centering
    \includegraphics[scale=0.5]{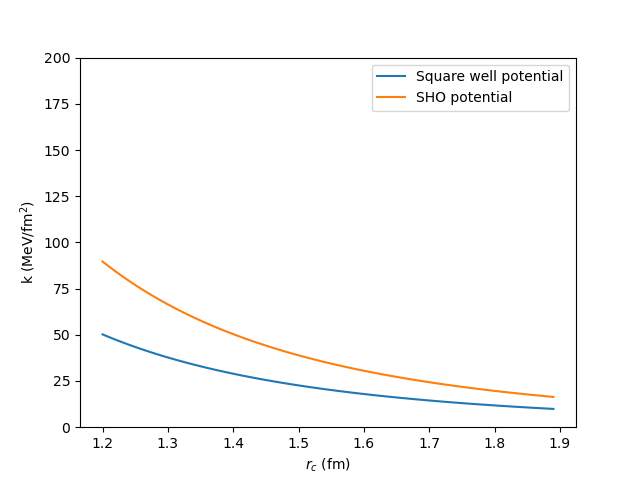}
    \caption{The critical points of the di-meson potentials [SHO: Eq.~(\ref{eq:kcritHO}), top; square well: Eq.~(\ref{eq:kcritSquare}), bottom] for the $\ccqq$ $1^{++}$ state.}
    \label{fig:ccqq1++HPvsSP}
\end{figure}

In Fig.~\ref{fig:ccqq1++HPvsSP} we plot the critical ($k$,$r_c$) pairs for both the SHO and square well potentials.  In this figure, we clearly see that the parameter space of $(k,r_c)$ pairs for which one can find a $m_{\de}$ value allowing a perfect fit to $M_X$ is \textit{smaller} when using a square well potential: The SHO potential permits the use of larger $(k,r_c)$ pairs before encountering the additional mass eigenvalue, its boundary condition being given by Eq.~(\ref{eq:kcritHO}).  Table~\ref{tab:V_0comparison} exhibits specific parameter sets for which one can explicitly compare the difference in $k_{\rm crit}$ (and thus $V_0$) values for corresponding $r_c$ values.  In fact, this difference is relatively consistent, with $k^{\textrm{square}}_{\textrm{crit}} / k^{\textrm{SHO}}_{\textrm{crit}} \simeq 0.58$ over the exhibited range of $r_c$.  Notably, while the square-well potential parameter space that gives an exact fit to Eqs.~(\ref{eq:ExpMasses}) is smaller, its critical points in Fig.~\ref{fig:ccqq1++HPvsSP} are still well fit by a power law:
\begin{equation} \label{eq:kcritSquare}
    k^{\textrm{square}}_{\rm crit}=0.096 \cdot r^{-3.576}_{c}.
\end{equation}

\begin{table*} 
\caption{Values of the model parameters at which the pure molecular state emerges, for both square well [Eq.~(\ref{eq:Vsquare})] and SHO [Eq.~(\ref{eq:VHO})] di-meson potentials.  $r_c$ is the cutoff range, the $k_{M \bbar M}$ values satisfying this condition are labeled in the text as $k_{\rm crit}$, $V_0 \equiv -V_{M \bbar M} ({\bf r} =0)$ is the central depth of each potential, and $k_{M \bbar M}$ for the square well potential is defined as $2V_0/r_c^2$.}
\setlength{\tabcolsep}{9pt}
\renewcommand{\arraystretch}{1.2}
\begin{tabular}{c | c  c | c  c }
 \hline\hline
 & \multicolumn{2}{c|}{Square well} & \multicolumn{2}{c}{SHO} \\
 $r_c$~(fm) & $k_{M \bbar M}$~(MeV/fm\textsuperscript{2}) & $V_0$~(MeV) & $k_{M \bbar M}$~(MeV/fm\textsuperscript{2}) & $V_0$~(MeV) \\
  \hline
 1.2 & 50 & 36 & 90 & 65 \\
 1.4 & 29 & 21 & 50 & 36 \\
 1.6 & 18 & 13 & 31 & 22 \\
 1.8 & 12 & \ 8 & 20 & 14
\\
 \hline \hline
\end{tabular}
\label{tab:V_0comparison}
\end{table*}

\begin{figure} 
    \centering
    \includegraphics[scale=0.5]{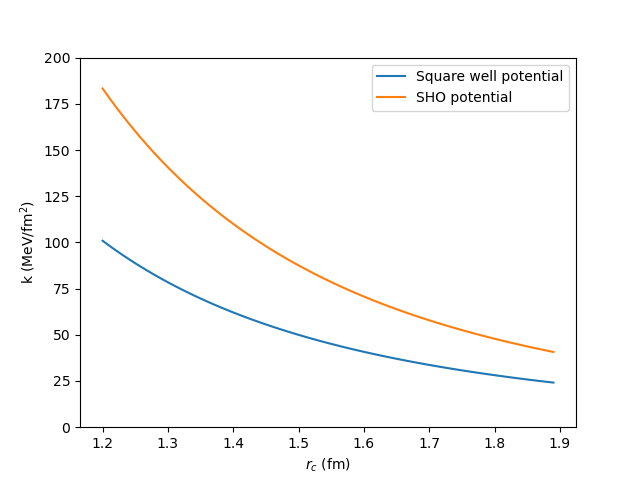}
    \caption{The same as Fig.~\ref{fig:ccqq1++HPvsSP} for the $\ccss$ $0^{++}$ state.}
    \label{fig:ccss0++HPvsSP}
\end{figure}

For the $0^{++}$ $c\bar c s\bar s$ state we find similar behavior, the primary difference being that this state is relatively more ``stable'': The critical points generally occur at \textit{larger} $(k,r_c)$ values.  This conclusion is apparent when one compares Fig.~\ref{fig:ccss0++HPvsSP} to Fig.~\ref{fig:ccqq1++HPvsSP}: The critical points of the SHO potential are well fit by
\begin{equation}
    k^{\textrm{SHO}}_{\rm crit}=0.335 \cdot r^{-3.312}_{c} ,
\end{equation}
and those for the square well potential satisfy
\begin{equation}
    k^{\textrm{square}}_{\rm crit}=0.179 \cdot r^{-3.149}_c.
\end{equation}

The smaller parameter space available for the square well potential before a pure molecular state appears is as one might expect: We are comparing potential wells with the same cutoff range $r_c$ and the same depth $V_0$.  The square well potential retains its full depth out to $r = r_c$, while the SHO potential approaches $V(r) = 0$ as $r \to r_c$, and thus its attractive di-meson interaction does not remain as strong over as large of a range as for the square-well case.

One additional direction within the $(k,r_c)$ space of great physical interest is that of large $r_c \in (10,20)$~fm.  As mentioned above, this range is much more in line with the expected size of an extremely near-threshold state like $\chi_{c1}(3872)$: The characteristic size $R$ of any sufficiently weakly bound 2-body quantum-mechanical bound state of reduced mass $\mu$ and (positive) binding energy $E$ is independent of the dynamics of the binding~\cite{Braaten:2003he}:
\begin{equation}
R \simeq \left( 2 \mu E \right)^{-1/2} .
\end{equation}
Using Eq.~(\ref{eq:Xbind}) for $-E$ produces the large, $O(10~\rm{fm})$ values of $R$ that motivate our expanded range for $r_c$.

\begin{figure} 
    \centering
    \includegraphics[scale=0.5]{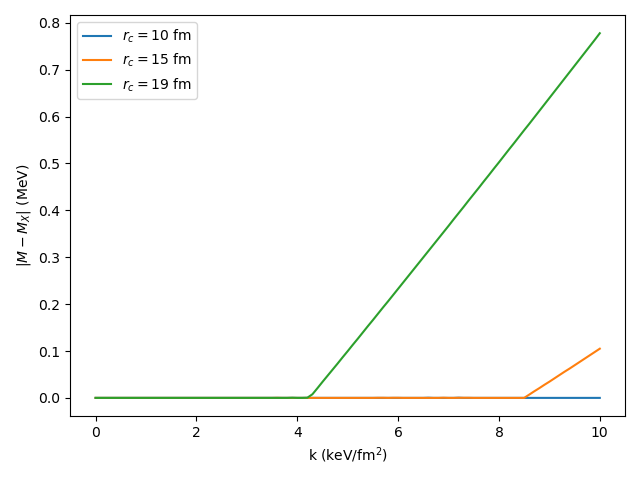}
    \caption{The $\ccqq$ $1^{++}$ mass difference $|M - M_X|$ as a function of $k$ for various values of $r_c$ (increasing from bottom to top) using the SHO potential Eq.~(\ref{eq:VHO}).}
    \label{fig:1++_k_massdiff_fine_Larger_c}
\end{figure}

\begin{figure} 
    \centering
    \includegraphics[scale=0.5]{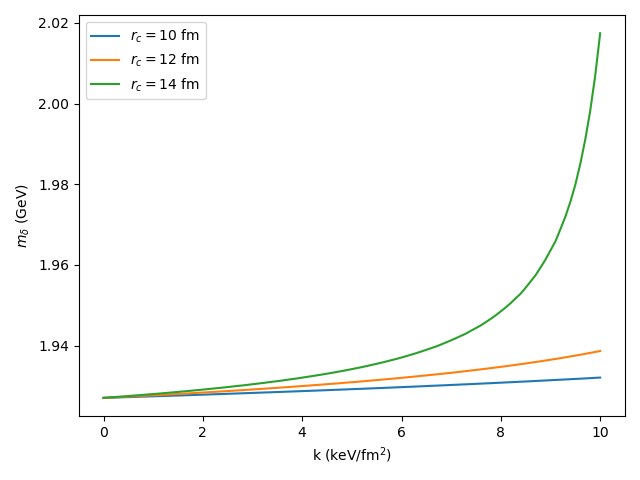}
    \caption{The $\ccqq$ diquark mass $m_{\de = (cq)}$ as a function of $k$ for various values of $r_c$ (increasing from bottom to top), using the SHO potential Eq.~(\ref{eq:VHO}).}
    \label{fig:1++_k_m_de_fine_Larger_c}
\end{figure}

\begin{figure} 
    \centering
    \includegraphics[scale=0.5]{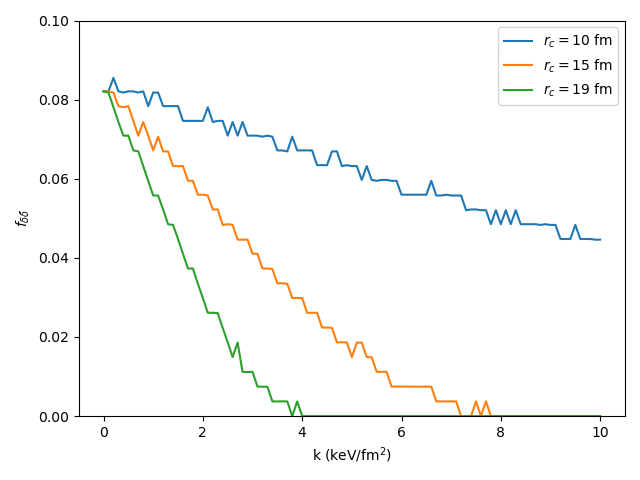}
    \caption{The $\de \bde$ fractional state content $f_{\de \bde}$ as a function of $k$ for various values of $r_c$ (increasing from top to bottom) for the $\ccqq ~ 1^{++}$ state, using the SHO potential Eq.~(\ref{eq:VHO}).  Discontinuities in the fit curve indicate that the extraction of this parameter lies at the edge of numerical precision for our algorithm.}
    \label{fig:1++_k_content_fine_SP2}
\end{figure}

In order to continue to probe the boundary region in which the pure molecular bound state arises, we must now reduce $k$ to a \textit{much} smaller range: $(0,10)~\rm keV/fm^2$\@.  We find that, while this region of parameter space does not generate potential wells nearly as deep (reaching a maximum of $V_0=1.8~\rm MeV$), it produces a significantly larger set of parameters in which the pure molecular bound state occurs.  Specifically, we see this state emerge as early as $r_c=15$~fm (Fig.~\ref{fig:1++_k_massdiff_fine_Larger_c}), which corresponds to $V_0 \in (0,1.1)~\rm MeV$ for the given $k$ range.  In contrast, despite yielding the same $V_0$ interval, the ranges $k \in (0,1) ~ \rm MeV/fm^2$ and $r_c=1.5~ \rm fm$ used above do not produce the pure molecular bound state at all.  We once again find that the critical values may be fit to a power law,
\begin{equation}
    k^{\textrm{SHO}}_{\rm crit}=9.72 \cdot 10^{-6} \, r^{-0.286}_{c},
\end{equation}
with $k$ still in units of GeV/fm$^2$ and $r_c$ in fm.  Apart from this key difference,  all results follow the same trends as those previously discussed. It remains true that fits to $m_\de (k)$ exhibit a strong dependence upon $r_c$ values in the region near the formation of the true di-meson molecular state [Fig.~\ref{fig:1++_k_m_de_fine_Larger_c}], and this dependence can even be successfully fit to an exponential over the finite region of large $r_c$ (10--19~fm) studied here, but the specific numerical dependence upon $r_c$ (values of the parameter $B$ given above) is not the same as for smaller $r_c$ ($\leq 1.9$~fm). In Fig.~\ref{fig:1++_k_massdiff_fine_Larger_c}, we see that the difference in mass eigenvalue between the experimental value of Eq.~(\ref{eq:ExpMasses}) and the molecular bound state for $r_c=19~\rm fm$ reaches a maximum value of $\simeq 800~ \rm keV$ by $k = 10$~keV/fm$^2$\@.  For $f_{\de \bde}$ [Fig.~\ref{fig:1++_k_content_fine_SP2}], we see the same behavior as before: decreasing to zero as the molecular state emerges, quadratic dependence on $k$ [Eq.~(\ref{eq:fdequad})], and its fit dependence upon $r_c$ being the same as discussed below Eq.~(\ref{eq:fdequad}).  Nevertheless, significant parameter space remains that produces substantial (several \%) $\de\bde$ state content. 

\section{Conclusions}\label{sec:concl}

In this paper we have addressed the question of how to generalize the diabatic dynamical diquark formalism, in which 4-quark exotic hadrons originate as diquark-antidiquark ($\de$-$\bde$) states, but whose properties become greatly modified if they lie in close proximity to di-meson thresholds.  The diabatic formalism is a rigorous, yet remarkably flexible extension of the Born-Oppenheimer approximation, and it allows one to incorporate a number of physically significant modifications.  Generalizing the approach of our previous works, in which the di-meson components prior to mixing are treated as being purely free two-particle states, here we explicitly include an attractive short-distance correction, {\it i.e.}, a di-meson potential.  Such a modification can truly be said to represent a full merging of diquark and meson molecular models for multiquark exotic hadrons.

This being an exploratory study, we obtain explicit results using two specific, simple potentials: a finite square well and a simple harmonic oscillator (SHO), both of which are truncated at a finite range $r_c$ and have depth $V_0$.  Any other potential that is finite and attractive at small di-meson separation and that asymptotes quickly to zero at large di-meson separation should give comparable results.  While the specific results we present (including studies of the functional dependence of the model under variation of its parameters) are quantitative only for these simple choices, the results are straightforward to describe in a qualitative manner: For substantial regions of $r_c$ and $V_0$ parameter space, typically bounded by $r_c \alt 2$~fm and $V_0 \alt 20$~MeV, the model for either potential produces a hybrid $\de\bde$/di-meson state that matches specific experimentally observed states [$\chi_{c1}(3872)$ and $\chi_{c0}(3915)$].  Specifically, allowing the diquark mass [$m_\de = m_{(cq)}$ or $m_{(cs)}$, respectively] to vary allows one to obtain a perfect fit to the measured state masses over a wide range of $r_c$ and $V_0$ values.  The explicit allowed parameter space for the square well potential is somewhat smaller than that for the SHO case, since it remains more strongly attractive over the entirety of the range $0 \leq r \leq r_c$.

If either the parameter $r_c$ or $V_0$ is permitted to become significantly larger than these ranges, then typically our numerical simulations become unstable.  To be explicit, it becomes extremely difficult to find a choice of model parameters for which the measured state mass eigenvalues can be successfully fit by the model.  At such a point, the model develops a new mass eigenvalue, corresponding to a purely molecular di-meson bound state.  The strongest evidence for this conclusion is the vanishing of the $\de \bde$ component of the eigenstate's wave function at the onset of this instability.  Such a result is not unexpected: Allowing a di-meson potential to be stronger than the nucleon-nucleon potential ought to produce di-meson molecular bound states, independent of the presence of a $\de$-$\bde$ attraction.

Lastly, we consider the effect of allowing $r_c$ to be very large [$O(15 \ {\rm fm})$] and $V_0$ to be very weak, as one might expect if treating the very near-threshold $\chi_{c1}(3872)$ as a pure $\bar D^0 D^{*0}$ molecule.  We find that, as one might expect, the values of $V_0$ for which pure molecular behavior arises become rather smaller, but otherwise substantial parameter ranges still occur in which the $\de\bde$ state content remains significant. 

With a method for a true unification of diquark-antidiquark and di-meson molecular models of 4-quark exotic hadrons now in hand, the next step involves determining the dependence of state mixing upon the specific spin and flavor states of the di-meson pair.  Significant experimental evidence for the presence of threshold effects in a variety of channels suggests that such phenomena are prominent, and the nature of their influence upon the formation of each exotic state is an important open question.

\begin{acknowledgments}
This work was supported by the National Science Foundation (NSF) under Grants No.\ PHY-2110278 and PHY-2405262.
\end{acknowledgments}


\bibliographystyle{apsrev4-2}
\bibliography{dimeson}

\begin{thebibliography}{34}%
\makeatletter
\providecommand \@ifxundefined [1]{%
 \@ifx{#1\undefined}
}%
\providecommand \@ifnum [1]{%
 \ifnum #1\expandafter \@firstoftwo
 \else \expandafter \@secondoftwo
 \fi
}%
\providecommand \@ifx [1]{%
 \ifx #1\expandafter \@firstoftwo
 \else \expandafter \@secondoftwo
 \fi
}%
\providecommand \natexlab [1]{#1}%
\providecommand \enquote  [1]{``#1''}%
\providecommand \bibnamefont  [1]{#1}%
\providecommand \bibfnamefont [1]{#1}%
\providecommand \citenamefont [1]{#1}%
\providecommand \href@noop [0]{\@secondoftwo}%
\providecommand \href [0]{\begingroup \@sanitize@url \@href}%
\providecommand \@href[1]{\@@startlink{#1}\@@href}%
\providecommand \@@href[1]{\endgroup#1\@@endlink}%
\providecommand \@sanitize@url [0]{\catcode `\\12\catcode `\$12\catcode `\&12\catcode `\#12\catcode `\^12\catcode `\_12\catcode `\%12\relax}%
\providecommand \@@startlink[1]{}%
\providecommand \@@endlink[0]{}%
\providecommand \url  [0]{\begingroup\@sanitize@url \@url }%
\providecommand \@url [1]{\endgroup\@href {#1}{\urlprefix }}%
\providecommand \urlprefix  [0]{URL }%
\providecommand \Eprint [0]{\href }%
\providecommand \doibase [0]{https://doi.org/}%
\providecommand \selectlanguage [0]{\@gobble}%
\providecommand \bibinfo  [0]{\@secondoftwo}%
\providecommand \bibfield  [0]{\@secondoftwo}%
\providecommand \translation [1]{[#1]}%
\providecommand \BibitemOpen [0]{}%
\providecommand \bibitemStop [0]{}%
\providecommand \bibitemNoStop [0]{.\EOS\space}%
\providecommand \EOS [0]{\spacefactor3000\relax}%
\providecommand \BibitemShut  [1]{\csname bibitem#1\endcsname}%
\let\auto@bib@innerbib\@empty
\bibitem [{\citenamefont {Choi}\ \emph {et~al.}(2003)\citenamefont {Choi} \emph {et~al.}}]{Choi:2003ue}%
  \BibitemOpen
  \bibfield  {author} {\bibinfo {author} {\bibfnamefont {S.}~\bibnamefont {Choi}} \emph {et~al.} (\bibinfo {collaboration} {Belle Collaboration}),\ }\href {https://doi.org/10.1103/PhysRevLett.91.262001} {\bibfield  {journal} {\bibinfo  {journal} {Phys.\ Rev.\ Lett.}\ }\textbf {\bibinfo {volume} {91}},\ \bibinfo {pages} {262001} (\bibinfo {year} {2003})},\ \Eprint {https://arxiv.org/abs/hep-ex/0309032} {arXiv:hep-ex/0309032} \BibitemShut {NoStop}%
\bibitem [{\citenamefont {Gell-Mann}(1964)}]{Gell-Mann:1964ewy}%
  \BibitemOpen
  \bibfield  {author} {\bibinfo {author} {\bibfnamefont {M.}~\bibnamefont {Gell-Mann}},\ }\href {https://doi.org/10.1016/S0031-9163(64)92001-3} {\bibfield  {journal} {\bibinfo  {journal} {Phys.\ Lett.}\ }\textbf {\bibinfo {volume} {{\bf 8}}},\ \bibinfo {pages} {214} (\bibinfo {year} {1964})}\BibitemShut {NoStop}%
\bibitem [{\citenamefont {Zweig}(1964)}]{Zweig:1964ruk}%
  \BibitemOpen
  \bibfield  {author} {\bibinfo {author} {\bibfnamefont {G.}~\bibnamefont {Zweig}},\ }\href {https://doi.org/10.17181/CERN-TH-401} {\bibinfo {title} {{\em An SU(3) Model for Strong Interaction Symmetry and Its Breaking. Version 1}}} (\bibinfo {year} {1964})\BibitemShut {NoStop}%
\bibitem [{\citenamefont {Zweig}()}]{Zweig:1964jf}%
  \BibitemOpen
  \bibfield  {author} {\bibinfo {author} {\bibfnamefont {G.}~\bibnamefont {Zweig}},\ }\bibinfo {title} {{An SU(3) Model for Strong Interaction Symmetry and Its Breaking. Version 2 (1964)}},\ in\ \href {https://doi.org/10.17181/CERN-TH-412} {\emph {\bibinfo {booktitle} {{Developments in the Quark Theory of Hadrons. Vol.\ 1. 1964--1978}}}},\ \bibinfo {editor} {edited by\ \bibinfo {editor} {\bibfnamefont {D.}~\bibnamefont {Lichtenberg}}\ and\ \bibinfo {editor} {\bibfnamefont {S.}~\bibnamefont {Rosen}}},\ p.~\bibinfo {pages} {22}\BibitemShut {NoStop}%
\bibitem [{\citenamefont {Voloshin}\ and\ \citenamefont {Okun}(1976)}]{Voloshin:1976ap}%
  \BibitemOpen
  \bibfield  {author} {\bibinfo {author} {\bibfnamefont {M.}~\bibnamefont {Voloshin}}\ and\ \bibinfo {author} {\bibfnamefont {L.}~\bibnamefont {Okun}},\ }\href@noop {} {\bibfield  {journal} {\bibinfo  {journal} {JETP Lett.}\ }\textbf {\bibinfo {volume} {{\bf 23}}},\ \bibinfo {pages} {333} (\bibinfo {year} {1976})}\BibitemShut {NoStop}%
\bibitem [{\citenamefont {De~Rujula}\ \emph {et~al.}(1975)\citenamefont {De~Rujula}, \citenamefont {Georgi},\ and\ \citenamefont {Glashow}}]{DeRujula:1975qlm}%
  \BibitemOpen
  \bibfield  {author} {\bibinfo {author} {\bibfnamefont {A.}~\bibnamefont {De~Rujula}}, \bibinfo {author} {\bibfnamefont {H.}~\bibnamefont {Georgi}},\ and\ \bibinfo {author} {\bibfnamefont {S.}~\bibnamefont {Glashow}},\ }\href {https://doi.org/10.1103/PhysRevD.12.147} {\bibfield  {journal} {\bibinfo  {journal} {Phys.\ Rev.\ D}\ }\textbf {\bibinfo {volume} {{\bf 12}}},\ \bibinfo {pages} {147} (\bibinfo {year} {1975})}\BibitemShut {NoStop}%
\bibitem [{\citenamefont {Navas}\ \emph {et~al.}(2024)\citenamefont {Navas} \emph {et~al.}}]{ParticleDataGroup:2024cfk}%
  \BibitemOpen
  \bibfield  {author} {\bibinfo {author} {\bibfnamefont {S.}~\bibnamefont {Navas}} \emph {et~al.} (\bibinfo {collaboration} {Particle Data Group}),\ }\href {https://doi.org/10.1103/PhysRevD.110.030001} {\bibfield  {journal} {\bibinfo  {journal} {Phys.\ Rev.\ D}\ }\textbf {\bibinfo {volume} {{\bf 110}}},\ \bibinfo {pages} {030001} (\bibinfo {year} {2024})}\BibitemShut {NoStop}%
\bibitem [{\citenamefont {Suzuki}(2005)}]{Suzuki:2005ha}%
  \BibitemOpen
  \bibfield  {author} {\bibinfo {author} {\bibfnamefont {M.}~\bibnamefont {Suzuki}},\ }\href {https://doi.org/10.1103/PhysRevD.72.114013} {\bibfield  {journal} {\bibinfo  {journal} {Phys.\ Rev.\ D}\ }\textbf {\bibinfo {volume} {{\bf 72}}},\ \bibinfo {pages} {114013} (\bibinfo {year} {2005})},\ \Eprint {https://arxiv.org/abs/hep-ph/0508258} {arXiv:hep-ph/0508258} \BibitemShut {NoStop}%
\bibitem [{\citenamefont {Lebed}(2015)}]{Lebed:2015tna}%
  \BibitemOpen
  \bibfield  {author} {\bibinfo {author} {\bibfnamefont {R.}~\bibnamefont {Lebed}},\ }\href {https://doi.org/10.1016/j.physletb.2015.08.032} {\bibfield  {journal} {\bibinfo  {journal} {Phys.\ Lett.\ B}\ }\textbf {\bibinfo {volume} {{\bf 749}}},\ \bibinfo {pages} {454} (\bibinfo {year} {2015})},\ \Eprint {https://arxiv.org/abs/1507.05867} {arXiv:1507.05867 [hep-ph]} \BibitemShut {NoStop}%
\bibitem [{\citenamefont {Brodsky}\ \emph {et~al.}(2014)\citenamefont {Brodsky}, \citenamefont {Hwang},\ and\ \citenamefont {Lebed}}]{Brodsky:2014xia}%
  \BibitemOpen
  \bibfield  {author} {\bibinfo {author} {\bibfnamefont {S.}~\bibnamefont {Brodsky}}, \bibinfo {author} {\bibfnamefont {D.}~\bibnamefont {Hwang}},\ and\ \bibinfo {author} {\bibfnamefont {R.}~\bibnamefont {Lebed}},\ }\href {https://doi.org/10.1103/PhysRevLett.113.112001} {\bibfield  {journal} {\bibinfo  {journal} {Phys.\ Rev.\ Lett.}\ }\textbf {\bibinfo {volume} {{\bf 113}}},\ \bibinfo {pages} {112001} (\bibinfo {year} {2014})},\ \Eprint {https://arxiv.org/abs/1406.7281} {arXiv:1406.7281 [hep-ph]} \BibitemShut {NoStop}%
\bibitem [{\citenamefont {Lebed}(2017)}]{Lebed:2017min}%
  \BibitemOpen
  \bibfield  {author} {\bibinfo {author} {\bibfnamefont {R.}~\bibnamefont {Lebed}},\ }\href {https://doi.org/10.1103/PhysRevD.96.116003} {\bibfield  {journal} {\bibinfo  {journal} {Phys.\ Rev.\ D}\ }\textbf {\bibinfo {volume} {{\bf 96}}},\ \bibinfo {pages} {116003} (\bibinfo {year} {2017})},\ \Eprint {https://arxiv.org/abs/1709.06097} {arXiv:1709.06097 [hep-ph]} \BibitemShut {NoStop}%
\bibitem [{\citenamefont {Giron}\ \emph {et~al.}(2019)\citenamefont {Giron}, \citenamefont {Lebed},\ and\ \citenamefont {Peterson}}]{Giron:2019bcs}%
  \BibitemOpen
  \bibfield  {author} {\bibinfo {author} {\bibfnamefont {J.}~\bibnamefont {Giron}}, \bibinfo {author} {\bibfnamefont {R.}~\bibnamefont {Lebed}},\ and\ \bibinfo {author} {\bibfnamefont {C.}~\bibnamefont {Peterson}},\ }\href {https://doi.org/10.1007/JHEP05(2019)061} {\bibfield  {journal} {\bibinfo  {journal} {J. High Energy Phys.}\ }\textbf {\bibinfo {volume} {{\bf 05}}},\ \bibinfo {pages} {061}},\ \Eprint {https://arxiv.org/abs/1903.04551} {arXiv:1903.04551 [hep-ph]} \BibitemShut {NoStop}%
\bibitem [{\citenamefont {Giron}\ \emph {et~al.}(2020)\citenamefont {Giron}, \citenamefont {Lebed},\ and\ \citenamefont {Peterson}}]{Giron:2019cfc}%
  \BibitemOpen
  \bibfield  {author} {\bibinfo {author} {\bibfnamefont {J.}~\bibnamefont {Giron}}, \bibinfo {author} {\bibfnamefont {R.}~\bibnamefont {Lebed}},\ and\ \bibinfo {author} {\bibfnamefont {C.}~\bibnamefont {Peterson}},\ }\href {https://doi.org/10.1007/JHEP01(2020)124} {\bibfield  {journal} {\bibinfo  {journal} {J. High Energy Phys.}\ }\textbf {\bibinfo {volume} {{\bf 01}}},\ \bibinfo {pages} {124}},\ \Eprint {https://arxiv.org/abs/1907.08546} {arXiv:1907.08546 [hep-ph]} \BibitemShut {NoStop}%
\bibitem [{\citenamefont {Giron}\ and\ \citenamefont {Lebed}(2020{\natexlab{a}})}]{Giron:2020fvd}%
  \BibitemOpen
  \bibfield  {author} {\bibinfo {author} {\bibfnamefont {J.}~\bibnamefont {Giron}}\ and\ \bibinfo {author} {\bibfnamefont {R.}~\bibnamefont {Lebed}},\ }\href {https://doi.org/10.1103/PhysRevD.101.074032} {\bibfield  {journal} {\bibinfo  {journal} {Phys.\ Rev.\ D}\ }\textbf {\bibinfo {volume} {{\bf 101}}},\ \bibinfo {pages} {074032} (\bibinfo {year} {2020}{\natexlab{a}})},\ \Eprint {https://arxiv.org/abs/2003.02802} {arXiv:2003.02802 [hep-ph]} \BibitemShut {NoStop}%
\bibitem [{\citenamefont {Giron}\ and\ \citenamefont {Lebed}(2020{\natexlab{b}})}]{Giron:2020qpb}%
  \BibitemOpen
  \bibfield  {author} {\bibinfo {author} {\bibfnamefont {J.}~\bibnamefont {Giron}}\ and\ \bibinfo {author} {\bibfnamefont {R.}~\bibnamefont {Lebed}},\ }\href {https://doi.org/10.1103/PhysRevD.102.014036} {\bibfield  {journal} {\bibinfo  {journal} {Phys.\ Rev.\ D}\ }\textbf {\bibinfo {volume} {{\bf 102}}},\ \bibinfo {pages} {014036} (\bibinfo {year} {2020}{\natexlab{b}})},\ \Eprint {https://arxiv.org/abs/2005.07100} {arXiv:2005.07100 [hep-ph]} \BibitemShut {NoStop}%
\bibitem [{\citenamefont {Giron}\ and\ \citenamefont {Lebed}(2020{\natexlab{c}})}]{Giron:2020wpx}%
  \BibitemOpen
  \bibfield  {author} {\bibinfo {author} {\bibfnamefont {J.}~\bibnamefont {Giron}}\ and\ \bibinfo {author} {\bibfnamefont {R.}~\bibnamefont {Lebed}},\ }\href {https://doi.org/10.1103/PhysRevD.102.074003} {\bibfield  {journal} {\bibinfo  {journal} {Phys.\ Rev.\ D}\ }\textbf {\bibinfo {volume} {{\bf 102}}},\ \bibinfo {pages} {074003} (\bibinfo {year} {2020}{\natexlab{c}})},\ \Eprint {https://arxiv.org/abs/2008.01631} {arXiv:2008.01631 [hep-ph]} \BibitemShut {NoStop}%
\bibitem [{\citenamefont {Giron}\ \emph {et~al.}(2021)\citenamefont {Giron}, \citenamefont {Lebed},\ and\ \citenamefont {Martinez}}]{Giron:2021sla}%
  \BibitemOpen
  \bibfield  {author} {\bibinfo {author} {\bibfnamefont {J.}~\bibnamefont {Giron}}, \bibinfo {author} {\bibfnamefont {R.}~\bibnamefont {Lebed}},\ and\ \bibinfo {author} {\bibfnamefont {S.}~\bibnamefont {Martinez}},\ }\href {https://doi.org/10.1103/PhysRevD.104.054001} {\bibfield  {journal} {\bibinfo  {journal} {Phys.\ Rev.\ D}\ }\textbf {\bibinfo {volume} {{\bf 104}}},\ \bibinfo {pages} {054001} (\bibinfo {year} {2021})},\ \Eprint {https://arxiv.org/abs/2106.05883} {arXiv:2106.05883 [hep-ph]} \BibitemShut {NoStop}%
\bibitem [{\citenamefont {Giron}\ and\ \citenamefont {Lebed}(2021)}]{Giron:2021fnl}%
  \BibitemOpen
  \bibfield  {author} {\bibinfo {author} {\bibfnamefont {J.}~\bibnamefont {Giron}}\ and\ \bibinfo {author} {\bibfnamefont {R.}~\bibnamefont {Lebed}},\ }\href {https://doi.org/10.1103/PhysRevD.104.114028} {\bibfield  {journal} {\bibinfo  {journal} {Phys.\ Rev.\ D}\ }\textbf {\bibinfo {volume} {{\bf 104}}},\ \bibinfo {pages} {114028} (\bibinfo {year} {2021})},\ \Eprint {https://arxiv.org/abs/2110.05557} {arXiv:2110.05557 [hep-ph]} \BibitemShut {NoStop}%
\bibitem [{\citenamefont {Lebed}\ and\ \citenamefont {Martinez}(2022)}]{Lebed:2022vks}%
  \BibitemOpen
  \bibfield  {author} {\bibinfo {author} {\bibfnamefont {R.}~\bibnamefont {Lebed}}\ and\ \bibinfo {author} {\bibfnamefont {S.}~\bibnamefont {Martinez}},\ }\href {https://doi.org/10.1103/PhysRevD.106.074007} {\bibfield  {journal} {\bibinfo  {journal} {Phys.\ Rev.\ D}\ }\textbf {\bibinfo {volume} {{\bf 106}}},\ \bibinfo {pages} {074007} (\bibinfo {year} {2022})},\ \Eprint {https://arxiv.org/abs/2207.01101} {arXiv:2207.01101 [hep-ph]} \BibitemShut {NoStop}%
\bibitem [{\citenamefont {Lebed}\ and\ \citenamefont {Martinez}(2023)}]{Lebed:2023kbm}%
  \BibitemOpen
  \bibfield  {author} {\bibinfo {author} {\bibfnamefont {R.}~\bibnamefont {Lebed}}\ and\ \bibinfo {author} {\bibfnamefont {S.}~\bibnamefont {Martinez}},\ }\href {https://doi.org/10.1103/PhysRevD.108.014013} {\bibfield  {journal} {\bibinfo  {journal} {Phys.\ Rev.\ D}\ }\textbf {\bibinfo {volume} {{\bf 108}}},\ \bibinfo {pages} {014013} (\bibinfo {year} {2023})},\ \Eprint {https://arxiv.org/abs/2305.09146} {arXiv:2305.09146 [hep-ph]} \BibitemShut {NoStop}%
\bibitem [{\citenamefont {Baer}(2006)}]{Baer:2006}%
  \BibitemOpen
  \bibfield  {author} {\bibinfo {author} {\bibfnamefont {M.}~\bibnamefont {Baer}},\ }\href@noop {} {\emph {\bibinfo {title} {\it Beyond Born-Oppenheimer: Electronic Nonadiabatic Coupling Terms and Conical Intersections}}}\ (\bibinfo  {publisher} {Wiley},\ \bibinfo {address} {New Jersey},\ \bibinfo {year} {2006})\BibitemShut {NoStop}%
\bibitem [{\citenamefont {Bruschini}\ and\ \citenamefont {Gonz\'{a}lez}(2020)}]{Bruschini:2020voj}%
  \BibitemOpen
  \bibfield  {author} {\bibinfo {author} {\bibfnamefont {R.}~\bibnamefont {Bruschini}}\ and\ \bibinfo {author} {\bibfnamefont {P.}~\bibnamefont {Gonz\'{a}lez}},\ }\href {https://doi.org/10.1103/PhysRevD.102.074002} {\bibfield  {journal} {\bibinfo  {journal} {Phys.\ Rev.\ D}\ }\textbf {\bibinfo {volume} {{\bf 102}}},\ \bibinfo {pages} {074002} (\bibinfo {year} {2020})},\ \Eprint {https://arxiv.org/abs/2007.07693} {arXiv:2007.07693 [hep-ph]} \BibitemShut {NoStop}%
\bibitem [{\citenamefont {Lebed}\ and\ \citenamefont {Martinez}(2024{\natexlab{a}})}]{Lebed:2024rsi}%
  \BibitemOpen
  \bibfield  {author} {\bibinfo {author} {\bibfnamefont {R.}~\bibnamefont {Lebed}}\ and\ \bibinfo {author} {\bibfnamefont {S.}~\bibnamefont {Martinez}},\ }\href {https://doi.org/10.1103/PhysRevD.110.074002} {\bibfield  {journal} {\bibinfo  {journal} {Phys.\ Rev\. D}\ }\textbf {\bibinfo {volume} {{\bf 110}}},\ \bibinfo {pages} {074002} (\bibinfo {year} {2024}{\natexlab{a}})},\ \Eprint {https://arxiv.org/abs/2404.15186} {arXiv:2404.15186 [hep-ph]} \BibitemShut {NoStop}%
\bibitem [{\citenamefont {Vijande}\ \emph {et~al.}(2007)\citenamefont {Vijande}, \citenamefont {Valcarce},\ and\ \citenamefont {Richard}}]{Vijande:2007ix}%
  \BibitemOpen
  \bibfield  {author} {\bibinfo {author} {\bibfnamefont {J.}~\bibnamefont {Vijande}}, \bibinfo {author} {\bibfnamefont {A.}~\bibnamefont {Valcarce}},\ and\ \bibinfo {author} {\bibfnamefont {J.-M.}\ \bibnamefont {Richard}},\ }\href {https://doi.org/10.1103/PhysRevD.76.114013} {\bibfield  {journal} {\bibinfo  {journal} {Phys.\ Rev.\ D}\ }\textbf {\bibinfo {volume} {{\bf 76}}},\ \bibinfo {pages} {114013} (\bibinfo {year} {2007})},\ \Eprint {https://arxiv.org/abs/0707.3996} {arXiv:0707.3996 [hep-ph]} \BibitemShut {NoStop}%
\bibitem [{\citenamefont {Dubynskiy}\ and\ \citenamefont {Voloshin}(2008)}]{Dubynskiy:2008mq}%
  \BibitemOpen
  \bibfield  {author} {\bibinfo {author} {\bibfnamefont {S.}~\bibnamefont {Dubynskiy}}\ and\ \bibinfo {author} {\bibfnamefont {M.}~\bibnamefont {Voloshin}},\ }\href {https://doi.org/10.1016/j.physletb.2008.07.086} {\bibfield  {journal} {\bibinfo  {journal} {Phys.\ Lett.\ B}\ }\textbf {\bibinfo {volume} {{\bf 666}}},\ \bibinfo {pages} {344} (\bibinfo {year} {2008})},\ \Eprint {https://arxiv.org/abs/0803.2224} {arXiv:0803.2224 [hep-ph]} \BibitemShut {NoStop}%
\bibitem [{\citenamefont {Lebed}\ and\ \citenamefont {Martinez}(2024{\natexlab{b}})}]{Lebed:2024zrp}%
  \BibitemOpen
  \bibfield  {author} {\bibinfo {author} {\bibfnamefont {R.}~\bibnamefont {Lebed}}\ and\ \bibinfo {author} {\bibfnamefont {S.}~\bibnamefont {Martinez}},\ }\href {https://doi.org/10.1103/PhysRevD.110.034033} {\bibfield  {journal} {\bibinfo  {journal} {Phys.\ Rev.\ D}\ }\textbf {\bibinfo {volume} {{\bf 110}}},\ \bibinfo {pages} {034033} (\bibinfo {year} {2024}{\natexlab{b}})},\ \Eprint {https://arxiv.org/abs/2406.08690} {arXiv:2406.08690 [hep-ph]} \BibitemShut {NoStop}%
\bibitem [{\citenamefont {Vijande}\ \emph {et~al.}(2009)\citenamefont {Vijande}, \citenamefont {Valcarce},\ and\ \citenamefont {Barnea}}]{Vijande:2009kj}%
  \BibitemOpen
  \bibfield  {author} {\bibinfo {author} {\bibfnamefont {J.}~\bibnamefont {Vijande}}, \bibinfo {author} {\bibfnamefont {A.}~\bibnamefont {Valcarce}},\ and\ \bibinfo {author} {\bibfnamefont {N.}~\bibnamefont {Barnea}},\ }\href {https://doi.org/10.1103/PhysRevD.79.074010} {\bibfield  {journal} {\bibinfo  {journal} {Phys. Rev. D}\ }\textbf {\bibinfo {volume} {79}},\ \bibinfo {pages} {074010} (\bibinfo {year} {2009})},\ \Eprint {https://arxiv.org/abs/0903.2949} {arXiv:0903.2949 [hep-ph]} \BibitemShut {NoStop}%
\bibitem [{\citenamefont {Swanson}(1992)}]{Swanson:1992ec}%
  \BibitemOpen
  \bibfield  {author} {\bibinfo {author} {\bibfnamefont {E.}~\bibnamefont {Swanson}},\ }\href {https://doi.org/10.1016/0003-4916(92)90327-I} {\bibfield  {journal} {\bibinfo  {journal} {Annals Phys.}\ }\textbf {\bibinfo {volume} {{\bf 220}}},\ \bibinfo {pages} {73} (\bibinfo {year} {1992})}\BibitemShut {NoStop}%
\bibitem [{Mor()}]{Morningstar:2019}%
  \BibitemOpen
  \href@noop {} {}\bibinfo {howpublished} {http://www.andrew.cmu.edu/user/cmorning/\\static\_potentials/SU3\_4D/greet.html}\BibitemShut {NoStop}%
\bibitem [{\citenamefont {Bruschini}(2023)}]{Bruschini:2023zkb}%
  \BibitemOpen
  \bibfield  {author} {\bibinfo {author} {\bibfnamefont {R.}~\bibnamefont {Bruschini}},\ }\href {https://doi.org/10.1007/JHEP08(2023)219} {\bibfield  {journal} {\bibinfo  {journal} {JHEP}\ }\textbf {\bibinfo {volume} {{\bf 08}}},\ \bibinfo {pages} {219}},\ \Eprint {https://arxiv.org/abs/2303.17533} {arXiv:2303.17533 [hep-ph]} \BibitemShut {NoStop}%
\bibitem [{\citenamefont {Kleiv}\ \emph {et~al.}(2013)\citenamefont {Kleiv}, \citenamefont {Steele}, \citenamefont {Zhang},\ and\ \citenamefont {Blokland}}]{Kleiv:2013dta}%
  \BibitemOpen
  \bibfield  {author} {\bibinfo {author} {\bibfnamefont {R.}~\bibnamefont {Kleiv}}, \bibinfo {author} {\bibfnamefont {T.}~\bibnamefont {Steele}}, \bibinfo {author} {\bibfnamefont {A.}~\bibnamefont {Zhang}},\ and\ \bibinfo {author} {\bibfnamefont {I.}~\bibnamefont {Blokland}},\ }\href {https://doi.org/10.1103/PhysRevD.87.125018} {\bibfield  {journal} {\bibinfo  {journal} {Phys.\ Rev.\ D}\ }\textbf {\bibinfo {volume} {{\bf 87}}},\ \bibinfo {pages} {125018} (\bibinfo {year} {2013})},\ \Eprint {https://arxiv.org/abs/1304.7816} {arXiv:1304.7816 [hep-ph]} \BibitemShut {NoStop}%
\bibitem [{\citenamefont {Acharya}\ \emph {et~al.}(2024)\citenamefont {Acharya} \emph {et~al.}}]{ALICE:2024bhk}%
  \BibitemOpen
  \bibfield  {author} {\bibinfo {author} {\bibfnamefont {S.}~\bibnamefont {Acharya}} \emph {et~al.} (\bibinfo {collaboration} {ALICE Collaboration}),\ }\href {https://doi.org/10.1103/PhysRevD.110.032004} {\bibfield  {journal} {\bibinfo  {journal} {Phys.\ Rev.\ D}\ }\textbf {\bibinfo {volume} {{\bf 110}}},\ \bibinfo {pages} {032004} (\bibinfo {year} {2024})},\ \Eprint {https://arxiv.org/abs/2401.13541} {arXiv:2401.13541 [nucl-ex]} \BibitemShut {NoStop}%
\bibitem [{\citenamefont {Lebed}\ and\ \citenamefont {Polosa}(2016)}]{Lebed:2016yvr}%
  \BibitemOpen
  \bibfield  {author} {\bibinfo {author} {\bibfnamefont {R.}~\bibnamefont {Lebed}}\ and\ \bibinfo {author} {\bibfnamefont {A.}~\bibnamefont {Polosa}},\ }\href {https://doi.org/10.1103/PhysRevD.93.094024} {\bibfield  {journal} {\bibinfo  {journal} {Phys.\ Rev.\ D}\ }\textbf {\bibinfo {volume} {{\bf 93}}},\ \bibinfo {pages} {094024} (\bibinfo {year} {2016})},\ \Eprint {https://arxiv.org/abs/1602.08421} {arXiv:1602.08421 [hep-ph]} \BibitemShut {NoStop}%
\bibitem [{\citenamefont {Braaten}\ and\ \citenamefont {Kusunoki}(2004)}]{Braaten:2003he}%
  \BibitemOpen
  \bibfield  {author} {\bibinfo {author} {\bibfnamefont {E.}~\bibnamefont {Braaten}}\ and\ \bibinfo {author} {\bibfnamefont {M.}~\bibnamefont {Kusunoki}},\ }\href {https://doi.org/10.1103/PhysRevD.69.074005} {\bibfield  {journal} {\bibinfo  {journal} {Phys.\ Rev.\ D}\ }\textbf {\bibinfo {volume} {{\bf 69}}},\ \bibinfo {pages} {074005} (\bibinfo {year} {2004})},\ \Eprint {https://arxiv.org/abs/hep-ph/0311147} {arXiv:hep-ph/0311147} \BibitemShut {NoStop}%
\end{thebibliography}%
\end{document}